\documentclass[aps,prd,preprint,tightenlines,superscriptaddress,
amsfonts,amssymb,amsmath,byrevtex,showpacs,nofootinbib]{revtex4-1}
\usepackage[polish,english]{babel}
\usepackage[utf8]{inputenc}
\usepackage[T1]{fontenc}
\usepackage{latexsym}
\usepackage{graphicx}
\usepackage{geometry}
\usepackage{epstopdf}
\usepackage{subfig}
\usepackage{color}

\usepackage{xcolor}
\usepackage{pdfsync}
\usepackage{fancyhdr}
\usepackage{makeidx}
\usepackage[breaklinks,colorlinks,backref]{hyperref}
\hypersetup{
    colorlinks, 
    linktoc=all, 
    linkcolor=red,  
  citecolor=hyptxt,
  urlcolor=blue}
  \hypersetup{
  citebordercolor=,
  filebordercolor=green,
  linkbordercolor=blue
}
\definecolor{hyptxt}{rgb}{0.7, 0.4, 0.9}
\usepackage{bibtopic}
\newcommand{\bea}{\begin{eqnarray}}
\newcommand{\eea}{\end{eqnarray}}

\newcommand{\RNumb}{\mathbb{R}}

\newcommand{\dR}{\mathbb R}

\newcommand{\UnitOp}{\hat{1\kern-4.75pt 1}} 
\newcommand{\MatUnit}{1\kern-3pt 1} 
\newcommand{\Aff}{\textrm{Aff}(\mathbb{R})} 
\newcommand{\Bra}[1]{\langle #1 \vert} 
\newcommand{\Ket}[1]{\vert #1 \rangle} 

\newcommand{\BraKet}[2]{\langle #1 \vert #2 \rangle} 

\newcommand{\Aver}[1]{\langle #1 \rangle} 
\newcommand{\sAver}[1]{\left\langle #1 \right\rangle} 

\newcommand{\Var}[1]{\mathrm{var}(#1)} 

\newcommand{\Komentarz}[1]{} 
\newcommand{\rmd}{{\mathrm{d}}}

\begin{document}

\title{Ascribing quantum system to Schwarzschild \\ spacetime
with naked singularity}

\author{Andrzej G\'{o}\'{z}d\'{z}}
\email{andrzej.gozdz@umcs.lublin.pl}
\affiliation{Institute of Physics, Maria Curie-Sk{\l}odowska
University, pl.  Marii Curie-Sk{\l}odowskiej 1, 20-031 Lublin, Poland}


\author{Aleksandra P\c{e}drak} \email{aleksandra.pedrak@ncbj.gov.pl}
\affiliation{Department of Fundamental Research, National Centre for Nuclear
  Research, Pasteura 7, 02-093 Warszawa, Poland}

\author{W{\l}odzimierz Piechocki} \email{wlodzimierz.piechocki@ncbj.gov.pl}
\affiliation{Department of Fundamental Research, National Centre for Nuclear
  Research, Pasteura 7, 02-093 Warszawa, Poland}

\date{\today}

\begin{abstract}
We quantize the Schwarzschild spacetime with naked singularity using the affine
coherent states quantization method.  The novelty of our approach is
quantization of both temporal and spatial coordinates.  Quantization smears the
gravitational singularity indicated by the Kretschmann invariant avoiding its
localization in the configuration space. This way we resolve the singularity
problem of considered spacetime at quantum level.
\end{abstract}


\maketitle


\section{Introduction}

One of the motivations of this paper is constructing the tools to be used
in the quantization of the Lemaître-Tolman-Bondi model of spacetime.
Another one is testing the idea of quantization of both temporal
and spatial variables of simple gravitational system to be used later in the
case of more sophisticated gravitational models.

The system we consider to be quantized is the celebrated Schwarzschild spacetime
\cite{Schw,Dro}.  We ascribe to this gravitational system a quantum system by
making use of the affine coherent states (ACS) approach that we have recently
used for the quantization of the Belinski-Khalatnikov-Lifshitz scenario with
generic cosmological singularity \cite{AWG,AW}.

To this end, we quantize not only spatial but also temporal coordinates.
Instead of phase space used in Hamiltonian formulations, we introduce the notion
of an extended configuration space including the time variable. This space is
used to quantize both elementary and composite observables.

As far as we are aware, our paper is the first one which proposes the
quantization of the temporal and spatial variables in general relativity. Quite
general rationale for such dealing is the following: the distinction between
space and time violates relativity; in particular, the general covariance of
arbitrary transformations of temporal and spatial coordinates.

By resolving the gravitational singularity problem, we mean showing that
quantization smears the singularity indicated by the Kretschmann scalar avoiding
its localization in the configuration space.

Recently, we have found that the ACS quantization depends on the choice of the
parametrization of the affine group \cite{AWT}. In this paper we present another
``parameter'' of the ACS method, unknown before, that is connected with the
freedom in the choice of the center of the affine group.

There are at least three goals of this paper: (i) presenting a powerful quantization method
especially suitable for quantization of gravitational systems, (ii) applying successfully  this method
to the resolution of gravitational singularity of an isolated object, and (iii) showing that treating
temporal and spatial coordinates on the same footing, supporting the covariance of general relativity,
enables the construction of consistent quantum theory.

The paper is organized as follows: In Sec. II we recall the known properties of
the Schwarzschild spacetime. Sec. III is devoted to the quantum theory. We
recall the formalism of the affine coherent states quantization method. Then, we
quantize the temporal and spatial coordinates which are elementary
observables. Quantization of the main observable, the Kretschmann scalar, is
carried out in Sec. IV.  It includes examination of the expectation value of
Kretschmann's operator.  We conclude in Sec.V.  Appendixes include some
practical rules concerning calculations of special expressions, eigensolutions
for elementary observables, expectation value of the Kretschmann operator within
some basis of the carrier space, and determination of some parameters used in
the paper.

\noindent In the following we choose $\;G = c =1\;$ except where otherwise
noted.

\section{Classical  model}

One of the simplest vacuum solutions to Einstein's equations, representing the
spherically symmetric black hole is the Schwarzschild spacetime.  The
Schwarzschild metric in the so-called Schwarzschild coordinates
$(t,r,\theta,\phi)\in \dR\times (0,\infty)\times S^2$ reads
\cite{black-bible,Piotr}:
\begin{equation}\label{LCh}
\rmd s^2 = -\left(1-\frac{r_s}{r}\right)\rmd t^2
+\left(1-\frac{r_s}{r}\right)^{-1}\rmd r^2
+r^2\left(\rmd \theta^2+\sin^2\theta \rmd \phi^2\right)\;,
\end{equation}
where $t$ is the time coordinate measured by a stationary clock located
infinitely far from black hole, $r$ is the radial coordinate measured as the
circumference (divided by $2 \pi$) of a sphere centered around the black hole,
$\theta$ and $\phi$ are angle coordinates of the sphere $S^2$, $r_s = 2M$
denotes the Schwarzschild radius defining the event horizon, and $M$ is the mass
parameter of the black hole.  It is commonly known that $r=r_s$ defines not
gravitational, but  a coordinate singularity.

The event horizon divides the Schwarzschild spacetime into the interior and
exterior regions of that black hole. The exterior metric, defined by
\eqref{LCh}, is static. In the interior region, the exterior spatial radial and
temporal coordinates exchange their character so that the metric coefficients
become time dependent \cite{Piotr}.  There exists the isometry of the interior
of the Schwarzschild black hole with the vacuum Kantowski-Sachs spacetime (see,
e.g. \cite{Edward}) which can be used for the quantization of the former.  We
make some remarks on that quantization in the concluding section.

In this paper we ascribe a quantum system to the Schwarzschild spacetime devoid
of the event horizon.  Such gravitational model is defined by the metric
\eqref{LCh} with $M < 0$, which is static for any $r > 0$ (see,
e.g. \cite{Piotr}).  This way we avoid the problem of bearing of the horizon on
the quantization which simplifies the latter.

To identify the curvature singularity, we cannot use the Ricci scalar and tensor
as these are vanishing for the vacuum solution.  However, another curvature
invariant, the Kretschmann scalar is non-zero and reads
\cite{black-bible,Piotr}:
\begin{equation}\label{Kret}
\mathcal{K} := R^{\alpha \beta \gamma \delta}R_{\alpha \beta \gamma \delta} =
\frac{48 M^2}{r^6}\;,
\end{equation}
so that it exhibits the gravitational singularity as $r\rightarrow 0$.

The Kretschmann invariant is the main observable to be examined at quantum
level.

\section{Quantum description}

The classical description of the model presented in the previous section
includes two elementary observables: time and radial coordinates.  The former is
timelike and the latter is spacelike.  In the standard quantization procedure,
the time variable may play the role of an evolution parameter as in the
Schr\"odinger equation. In what follows, we quantize both the temporal and
spatial coordinates.
For both variables the time $t$ and the radial coordinates $r$, we construct
  their quantum counterparts. Both quantum observables (operators) we treat on
  the same footing. It means that time is no longer a parameter, but similarly as
  the radial coordinate, a quantum observable represented by an appropriate
  operator obtained by a quantization procedure. In the following, as it was
  mentioned earlier, we are using the affine coherent states quantization
  (ACS). As we will see later the ACS quantization leads to the operators
  $\hat{t}$ and $\hat{r}$ which, in general, do not commute. Due to the
  Heisenberg uncertainty principle they cannot be considered as a compatible
  pair of quantum observables. In addition, because of this property, one cannot
  construct the common eigenstates of both observables, which would represent
  spacetime position states. In such case, the most appropriate candidates for
  the spacetime position states are the coheret states. The coherent states
  furnish a set of non--orthogonal states. It means that, in general, the
  spacetime position states are always connected by a non--zero transition
  amplitudes.  They cannot be considered as a set of independent alternatives as
  it is in the case of common eigenstates of commuting self-adjoint
  operators. 

In this paper we want to check if introducing of time as quantum observables can
help to resolve at the quantum level the main problem of general relativity,
which is the existence of solutions with gravitational singularities.  To begin
with, we address the singularity problem of the simplest solution to Einstein's
gravity, but we plan to apply this approach to more advanced singular solutions
within general relativity.

\subsection{Affine coherent states quantization}

The covariance of general relativity requires to treat both variables $t$
  and $r$ on the same footing in both the classical and quantum descriptions.
  To fulfill this condition we begin with introducing the notion of the extended
  configuration space $T$ of our system by including time as next coordinate
  variable required in description of this quantum system.  It is defined as
follows
\begin{equation}
\label{t1}
  T = \{(t,r) ~|~ (t,r) \in \RNumb \times \RNumb_+\},~~~~ \RNumb_+ = (0,+\infty)
  \, ,
\end{equation}
where $t$ and $r$ are the time and the radial coordinates, respectively, which
occur in the line element \eqref{LCh}.
The corresponding operators $\hat{t}$ and
$\hat{r}$ are constructed in the subsection B by the ACS quantization
procedure.

As usually in quantum mechanics, to define a quantum observable one needs to
determine an operational procedure which connects this observable with its
quantum description. In the case of $t$ and $r$ one needs to measure time and
spacial distance. To relate the values of measured time and radial variables
to our states and operators we introduce the consistency conditions
\eqref{eq:TSelfConsCond} and \eqref{eq:RSelfConsCond}. They  represent compatibility
of expectation values of the time and position operators, within the coherent
states, with measured values.

The other space variables $\theta$ and $\phi$ of \eqref{LCh}, used to implement
the spherical symmetry of considered spacetime, do not enter the definition of
$T$ as the main observable to be quantized, the Kretschmann scalar, does not
depend on these variables. 
In the following we sketch the basic facts about affine quantization required in
  further considerations. The most important formula in this subsection is the
  expression \eqref{eq:3AffQuantization} for quantization of any arbitrary
  classical mechanics function defined on the configuration
  space of a given physical system.

Since the configuration space is a half-plane, every point $(t,r) \in T$ can be
uniquely identified with the corresponding element $g(\chi_1(t,r),\chi_2(t,r))$
of the affine group $\Aff$, where $\chi(t,r)=(\chi_1(t,r),\chi_2(t,r))$ is a
one-to-one mapping between $T$ and any arbitrary chosen fixed parametrization of
$\Aff$.

As the standard parametrization of the affine group (see \cite{AWT} for more
details) we assume the parametrization $(p,q) \in \RNumb \times \RNumb_+ $ which
obey the following multiplication law
\begin{equation}\label{t5}
 g(p_1, q_1) \cdot g(p_2, q_2) := g(p_1 + q_1 p_2, q_1 q_2) \in \Aff \, ,
\end{equation}
and the left invariant measure on this group is defined as
\begin{equation}\label{eq:gParHaar}
 d\mu(p,q)= dp \frac{dq}{q^2} \, .
\end{equation}
The corresponding left invariant integration over the affine group is given by
\begin{equation}\label{t7}
  \int_{\Aff} d\mu (p,q) :=
  \frac{1}{2\pi}\int_{-\infty}^\infty dp \int_0^\infty dq /q^2 \, .
\end{equation}
It enables defining the Hilbert space of functions on the affine group
$\mathcal{H}_g := L^2( \Aff, d\mu (g))$, where $g=g(p,q) \in \Aff$.

Because $(p,q)=\chi(t,r)$ is a one-to-one function, the coordinates
$(t,r) \in T$ also parameterize the affine group $\Aff$, i.e., we have the
mapping $(t,r) \to g(\chi_1(t,r),\chi_2(t,r))$.  The corresponding measure
(not necessarily invariant) in the $(t,r)$ parametrization reads
\begin{equation}
\label{eq:hParHaar}
d\mu(p,q)= \sigma(t,r) dt dr=
\begin{vmatrix}
    \frac{\partial\chi_1}{\partial t} &
    \frac{\partial\chi_2}{\partial t} \\
    \frac{\partial\chi_1}{\partial r} &
    \frac{\partial\chi_2}{\partial r}
\end{vmatrix} dt dr  =: d\lambda(t,r) \, .
\end{equation}

It is known (see \cite{AWT} for more details) that the affine group has two
(inequivalent) irreducible unitary representations defined in the Hilbert space
$\mathcal{H}_x:= L^2 (\RNumb_+, d\nu(x))$, where $d\nu(x):= dx/x$.  We choose
the one defined as follows
\begin{equation}
\label{t6}
U (p,q)\Psi (x) := e^{ipx} \Psi (qx) \, ,
\end{equation}
where $\Psi (x) \in \mathcal{H}_x$.

The  carrier space  $\mathcal{H}_x$ is known  to have the basis \cite{GM}
\begin{equation}
\label{aa1}
e^{(\alpha)}_n (x) =
\sqrt{\frac{n!}{(n + \alpha)!}}\,e^{-x/2} x^{(1 +\alpha)/2}\,L_n^{(\alpha)}(x),
\end{equation}
where $L_n^{(\alpha)}$ is the Laguerre polynomial, $\alpha > -1$, and $(n +
\alpha)! = \Gamma (n + \alpha + 1)$.  One can verify that $\int_0^\infty
e^{(\alpha)}_n (x) e^{(\alpha)}_m (x) d\nu(x)= \delta_{n m}$ so that
$e^{(\alpha)}_n (x)$ is an orthonormal basis.

The coherent states in the  standard parametrization of the affine group,
$\BraKet{x}{g(p,q)} \in \mathcal{H}_x$, are defined as
follows
\begin{equation}\label{t8}
\BraKet{x}{g(p,q)} = U (p,q)\Phi_0 (x)= e^{ipx} \Phi_0 (qx) \, ,
\end{equation}
where $\Phi_0(x) \in \mathcal{H}_x$ is the so-called fiducial vector.  It is a
sort of a free ``parameter'' of the ACS quantization. First of all, it should be
normalized so that we should have
\begin{equation}\label{fid}
  \BraKet{\Phi_0}{\Phi_0} := \int_0^{\infty} d\nu (x)
  \BraKet{\Phi_0}{x} \BraKet{x}{\Phi_0}
  =  \int_0^{\infty} d\nu (x) |\Phi_0(x)|^2 = 1 \, ,
\end{equation}
where we have used the formula  \cite{AWT}
\begin{equation}\label{xxx}
\int_0^{\infty} d\nu (x) \Ket{x}\Bra{x}  =  \UnitOp \, ,
\end{equation}
which applies to $\mathcal{H}_x$.

The resolutions of the identity $\UnitOp$ in the Hilbert space $\mathcal{H}_x$,
in terms of the coherent states, reads \cite{AWT}
\begin{equation}\label{t9}
  \frac{1}{A_{\Phi_0}}\,\int_{\Aff} d\mu (p,q)~
  \Ket{g(p,q)} \Bra{g(p,q)} = \UnitOp  \, ,
\end{equation}
where
\begin{equation}\label{t10}
 A_{\Phi_0} := \int_0^\infty \frac{dx}{x^2} |\Phi_0(x)|^2  < \infty \, ,
\end{equation}
which defines another condition to be imposed on the fiducial vector
$\Phi_0(x)$.

Using \eqref{t9} we can (formally) map any observable $f: T \rightarrow \RNumb$
into a {\it symmetric} operator
$\hat{f}: \mathcal{H}_x\rightarrow \mathcal{H}_x$ as follows (see,
App.~(\ref{app:0}) and \cite{AWG,AWT} for more details)
\begin{equation}
\label{t11}
  \hat{f} :=
  \frac{1}{A_{\Phi_0}} \int_{\Aff} d\mu(p,q) \Ket{g(p,q)} f(p,q)\Bra{g(p,q)} \, .
\end{equation}

However, as it was shown in the paper \cite{AWT},  the affine quantization is
dependent on the parametrization of the affine group $\Aff$ and it has to be
considered also as a kind of a free ``parameter'' in the ACS quantization.

A fundamental expression in the ACS quantization is a non-orthogonal
decomposition of unity constructed from the coherent states
$\Ket{h(t,r)}=\Ket{g(\chi(t,r)}$:
\begin{equation}
 \label{eq:ResUnityParh}
  \frac{1}{A_{\Phi_0}}\,\int_{\Aff} d\lambda(t,r)~
  \Ket{h(t,r)} \Bra{h(t,r)}
 = \frac{1}{A_{\Phi_0}}\,\int_{\Aff} d\mu(p,q)~
  \Ket{g(p,q)} \Bra{g(p,q)} =\UnitOp  \, .
\end{equation}

The affine group manifold itself is a homogenous space. All points in this
manifold are equivalent to each other.  This means that from the physical
point of view we have an additional freedom in mapping of the configuration
space onto the group manifold $g(\chi(,)):T \to \Aff$.  More precisely,
using the ACS quantization it is usually assumed
that the element $(t_0=0,r_0=1) \in T$ of the configuration space is mapped
onto the unit element $g(0,1)$ of the affine group $\Aff$. Because of the
homogeneity of the group manifold, this assignment is in fact arbitrary. Every
choice of the mapping $g(\chi(t,r)) \in \Aff$ fixes in some way a relative
position between configuration space and the affine group manifold by the
relation $T \ni (t_0=0,r_0=1) \to g(a,b)=g(\chi(0,1))$. It defines the point
$g(a,b) \in\Aff$ which we call the ``center'' of the group manifold associated
to this configuration space. In the standard parametrization, where the
transformation $\chi$ is the identity transformation, the center is
identified with the unity $g(0,1)$ of the affine group.

To use this freedom one can check that the resolution of unity is invariant
with respect to any arbitrary left shift operation of the affine group
manifold:
\begin{equation}
 \label{eq:LeftShResUnity}
  \frac{1}{A_{\Phi_0}}\,\int_{\Aff} d\mu(p,q)~
  \Ket{g(a,b) \cdot g(p,q)} \Bra{g(a,b) \cdot g(p,q)} =\UnitOp  \, ,
\end{equation}
however, it is not invariant with respect to the right shift operation:
\begin{equation}
 \label{eq:LeftShResUnity}
  \frac{1}{A_{\Phi_0}}\,\int_{\Aff} d\mu(p,q)~
  \Ket{g(p,q) \cdot g(a,b)} \Bra{g(p,q) \cdot g(a,b)}
  = \Delta(g(a,b)^{-1})\UnitOp
  \, .
\end{equation}
In general, the function $\Delta(g)$ is the Haar modulus of the Lie group $G$
defined as
\begin{equation}
\label{eq:HaarMod}
\int_{G} d\mu(g) f(g \cdot h) =: \Delta(h^{-1}) \int_{G} d\mu(g) f(g)\, ,
\end{equation}
where $d\mu(g)$ denotes the left invariant measure on $G$.  Note that
the right shift of the unity resolution is still proportional to
resolution of unity. The right shift translates the ``center'' of the affine
group manifold to the new point $g(a,b)$.

Summing up, in the ACS quantization, which is a deformation of the resolution of
the unit operator, we have three free ``parameters'': choice of a fiducial
vector, choice of an affine group parametrization, and choice of a center of
the group manifold.

In fact, any choice of the center can be done by an appropriate choice of
the mapping $(p,q)=\chi(t,r)$. However, from technical point of view it is
useful to distinguish both operations: choice of the group parametrization and
a choice of the appropriate center, because the left $g'=g \cdot \tilde{g}$
and right $g''=\tilde{g}\cdot g$ shift of the element $g \in \Aff$ on the
group manifold commutes, i.e., both operations are independent. This is useful
property in calculations with invariant measure.

Using this freedom, the quantization process \eqref{t11} can be now generalized
to a deformation of the resolution of unity rewritten in a general
parametrization with the additional right shift which fixes the center of the
mapping between the configuration space and $\Aff$. Again introducing the
shortcut $\Ket{h(t,r)}=\Ket{g(\chi(t,r))}$ the required resolution of
unity read
\begin{equation}
\label{eq:GenResUnity}
\frac{\Delta(h(a',b'))}{A_{\Phi_0}}
\int_{\Aff} d\lambda(t,r)
\Ket{h(t,r) \cdot h(a',b')} \Bra{h(t,r) \cdot h(a',b')} =\UnitOp \, .
\end{equation}

Now, the ACS quantization of any function $f(t,r)$ on the configuration space is
defined as
\begin{equation}
\label{eq:AffQuantization}
\hat{f}=  \frac{\Delta(h(a',b'))}{A_{\Phi_0}}
\int_{\Aff} d\lambda(t,r)
\Ket{h(t,r) \cdot h(a',b')} f(t,r) \Bra{h(t,r) \cdot h(a',b')}  \, ,
\end{equation}
where the shift of the group manifold center is given by
$h(a',b')=g(\chi(a',b'))=:g(a,b) \in \Aff$.

It is useful to rewrite this formula in the form of our standard affine group
parametrization
\begin{equation}
\label{eq:2AffQuantization}
\hat{f}=  \frac{\Delta(h(a',b'))}{A_{\Phi_0}}
\int_{\Aff} d\lambda(t,r)
\Ket{g(\chi(t,r))\cdot g(a,b)} f(t,r) \Bra{g(\chi(t,r))\cdot g(a,b)} \, .
\end{equation}
After change of variables under integral
$p=\chi_1(t,r)$ and $q=\chi_2(t,r)$, and performing the right shift operation
\eqref{eq:HaarMod} in the coherent states, one gets the final expression for
quantization of the function $f(t,r)$:
\begin{equation}
\label{eq:3AffQuantization}
\hat{f}= \frac{1}{A_{\Phi_0}} \int_{\Aff} d\mu(p,q)
\Ket{g(p,q)}
f\left(\chi^{-1}\left(p-\frac{a}{b}q,\frac{q}{b} \right)\right)
\Bra{g(p,q)} \, .
\end{equation}
%

\subsection{Quantization of elementary observables}

The formula \eqref{eq:3AffQuantization} allows to quantize almost any real
function on the configuration space, $T$, giving the corresponding operator.
The most elementary observables are time and radial coordinates
\begin{eqnarray}
\label{eq:trOperators}
&& \hat{t}=\frac{1}{A_\Phi} \int_{\Aff} d\mu(p,q)
\Ket{g(p,q)} \chi^{-1}_1\left(p-\frac{a}{b}q,\frac{q}{b} \right)
\Bra{g(p,q)}\, ,  \\
&& \label{drugie}\hat{r}= \frac{1}{A_\Phi} \int_{\Aff} d\mu(p,q)
\Ket{g(p,q)} \chi^{-1}_2\left(p-\frac{a}{b}q,\frac{q}{b} \right)
\Bra{g(p,q)}\, .
\end{eqnarray}
They are required for description of the Schwarzschild spacetime.

As it was mentioned earlier in the subsection A, we have to relate the values of
measured time and coordinate with our quantum description. For this purpose
we have to choose the group parametrization and the group manifold
center $g(a,b)$ to fulfil the following consistency conditions:
\begin{eqnarray}
&&  \Aver{\hat{t};h(t,r)} = t
\label{eq:TSelfConsCond} \, , \\
&& \Aver{\hat{r};h(t,r)} = r\, ,
\label{eq:RSelfConsCond}
\end{eqnarray}
where $\Aver{\hat{A};\psi} := \Bra{\psi} \hat{A} \Ket{\psi}$ denotes expectation
value of the observable $\hat{A}$ in the state labelled by $\psi$.  These
condition relates the measured time and radial coordinate to the corresponding
quantum observables and states. 

It turns out that we do not need to reparameterize our group to fulfil required
conditions \eqref{eq:TSelfConsCond} and \eqref{eq:RSelfConsCond}.  We only have
to choose properly the group manifold center parameters $g(a,b)$ in the standard
parametrization. In general, these parameters are dependent on the
choice of the fiducial vector.

In the following we get two useful expressions for expectation values within
the coherent states $\Ket{g(t,r)}$, which can be easily obtained by
applying invariance of the Haar measure.

For any arbitrary operator \eqref{eq:3AffQuantization} quantized by means of the
affine group we get
\begin{eqnarray}
\label{eq:fAverGpar}
&&\langle \hat{f}; g(t,r)\rangle =
\frac{1}{A_{\Phi_0}} \int_{\Aff} d\mu(p,q)
\BraKet{g(t,r)}{g(p,q)} f\left(p-\frac{a}{b}q,\frac{q}{b} \right)
\BraKet{g(p,q)}{g(t,r)} \nonumber \\
&& = \frac{1}{A_{\Phi_0}} \int_{\Aff} d\mu(p,q) |\BraKet{g(0,1)}{g(p,q)}|^2
f\left(t+\left(p-\frac{a}{b}q\right)r,\frac{q}{b}r\right) \, ,
\end{eqnarray}
and for any product of two such operators we obtain
\begin{eqnarray}
\label{eq:ffAverGpar}
&&\langle \hat{f}_1 \hat{f}_2; g(t,r)\rangle =
\frac{1}{A_{\Phi_)}} \int_{\Aff} d\mu(p_1,q_1)
\frac{1}{A_{\Phi_0}} \int_{\Aff} d\mu(p_2,q_2)  \nonumber \\
&& f_1\left(t+(p_1-\frac{a}{b}q_1)r,\frac{q_1}{b}r\right)
f_2\left(t+\left(p_2-\frac{a}{b}q_2\right)r,\frac{q_2}{b}r\right) \nonumber \\
&& \BraKet{g(0,1)}{g(p_1,q_1)} \BraKet{g(p_1,q_1)}{g(p_2,q_2)}
\BraKet{g(p_2,q_2)}{g(0,1)} \, .
\end{eqnarray}

Using the formula \eqref{eq:fAverGpar}, the expectation value for the time
observable can be written in the following form
\begin{equation}
\label{eq:ExpValTime}
\Aver{\hat{t};g(t,r)}
=
\frac{1}{A_\Phi} \int_{\Aff} d\mu(p,q)
|\BraKet{g(0,1)}{g(p,q)}|^2 \left(t+r \left(p-\frac{a}{b}q\right)\right)  \, .\\
\end{equation}
After integration over $p$ and $q$ one gets
\begin{equation}
\label{eq:2ExpValTime}
\Aver{\hat{t};g(t,r)}
=t + \left(\Aver{\check{p}}_0-\frac{a}{b}\Aver{\check{q}}_0\right)r \, ,
\end{equation}
where we introduce the abbreviations:
\begin{equation}
\label{eq:gOperParam}
\check{f}:= \frac{1}{A_\Phi} \int_{\Aff} d\mu(p,q)
\Ket{g(p,q)} f(p,q) \Bra{g(p,q)} \, ,
\end{equation}
and
\begin{equation}
\label{eq:egAver}
\Aver{\check{f}}_0 := \frac{1}{A_\Phi} \int_{\Aff} d\mu(p,q)
|\BraKet{g(0,1)}{g(p,q)}|^2 f(p,q)= \Bra{g(0,1)} \check{f} \Ket{g(0,1)} \, .
\end{equation}
Thus, $\Aver{\check{f}}_0$ denotes the expectation value of the operator
$\check{f}$ in the fixed coherent state $\Ket{g(0,1)}$ corresponding to the
unity of the affine group.  In the case of more complicated expressions like
$f_1^nf_2^m$ instead of the notation \eqref{eq:gOperParam}, where the check
symbol is over the expression, we write $(f_1^nf_2^m)\check{\phantom{w}}$.

Similarly, we obtain
\begin{equation}
\label{eq:ExpValRadius}
\Aver{\hat{r};g(t,r)} = \frac{\Aver{\check{q}}_0}{b} r  \, .
\end{equation}
Assuming $\Aver{\check{p}}_0-\frac{a}{b}\Aver{\check{q}}_0=0$ and
$b=\Aver{\check{q}}_0$, i.e. $a=\Aver{\check{p}}_0$, the self consistency
conditions \eqref{eq:TSelfConsCond} and \eqref{eq:RSelfConsCond} become
fulfilled.\\

An important property of any quantum observable $\hat{A}$ is its {\it
variance}. The variance determines the value of  smearing of a quantum
observable. This influences behaviour of a given physical system substantially.
In the quantum state labelled by $\psi$ the variance is defined as follows
\begin{equation}
\label{eq:VarDef}
\Var{\hat{A};\psi}
:= \Aver{(\hat{A} - \Aver{\hat{A};\psi})^2;\psi}
=\Aver{\hat{A}^2;\psi} - \Aver{\hat{A};\psi}^2 \, .
\end{equation}
Formally, the variance is the stochastic deviation from the expectation value of
the observable $\hat{A}$.\\

 Suppose the operator $\hat{A}$ is essentially self-adjoint on some dense
 subspace $\mathcal{S}$ of the Hilbert space $\mathcal{H}_x$.  For every
 quantum  state $\psi \in \mathcal{S}$ of a physical system which belongs to
 the domain of the operator $A$ one can check that
\begin{equation}\label{lem}
\Big(\Var{\hat{A};\psi} = 0 \Big)\Longleftrightarrow \Big(\hat{A} \psi
= \lambda \psi  ,~~~\lambda \in \dR \Big) \,,
\end{equation}
i.e., the variance of the operator $\hat{A}$ is equal to 0, if and only if, the
quantum system is in an eigenstate of the operator $\hat{A}$. Then the
corresponding observable is not smeared.

The statement \eqref{lem} is implied by properties of the scalar product, norm
and the operator itself:
\[  \Var{\hat{A};\psi} = \langle(\hat{A} -
\langle\hat{A};\psi\rangle)\psi\,|\,(\hat{A} -
\langle\hat{A};\psi\rangle)\psi\rangle
= \|(\hat{A} - \langle\hat{A};\psi\rangle)\psi\|^2 \, . \]
Thus,
\[ \Big( \Var{\hat{A};\psi} = 0\Big) \Rightarrow \Big((\hat{A} -
\langle\hat{A};\psi\rangle)\psi
= 0 \Big) \Rightarrow \Big(\hat{A}\psi
= \langle\hat{A};\psi\rangle \psi \Big)\, .\]
The latter equality  means that $\langle\hat{A};\psi\rangle$ is the eigenvalue
of $\hat{A}$ corresponding to the eigenstate $\psi$.

\noindent On the other hand, if $\hat{A} \psi = \lambda \psi$, we have
\[ \Var{\hat{A};\psi} = \Aver{\hat{A}^2;\psi} - \Aver{\hat{A};\psi}^2 =
\lambda^2 \langle\psi | \psi \rangle - \lambda^2 \langle\psi | \psi \rangle^2 =0
\, , \]
as $\psi$ is a normalized vector. This completes the verification of the
validity of \eqref{lem}.

The variances of the operators $\hat{t}$ and $\hat{r}$ in the coherent states
$\Ket{g(t,r)}$ can be directly calculated.
They describe the smearing of both observables. The behaviour of variances and
  expectation values for time and radial coordinate allows to determine if they
  behave similarly to their classical counterparts or not.

Because of the self-consistency condition the only unknown components are
$\Aver{\hat{t}^2;g(t,r)}$ and $\Aver{\hat{r}^2;g(t,r)}$. Using the formula
\eqref{eq:ffAverGpar}
\begin{equation}
\label{eq:TVarOperator}
\Aver{\hat{t}^2;g(t,r)}
= t^2 +
2 \sAver{\check{p} -\frac{\Aver{\check{p}}_0}{\Aver{\check{q}}_0}\check{q}}_0tr
+ \sigma_t r^2 \, ,
\end{equation}
where
\begin{equation}
\label{eq:3TVarOperator}
\sigma_t :=   \sAver{\check{p}^2
  -\frac{\Aver{\check{p}}_0}{\Aver{\check{q}}_0}
  (\check{p}\check{q}+\check{q}\check{p})
  + \left(\frac{\Aver{\check{p}}_0}{\Aver{\check{q}}_0}\right)^2
  \check{q}^2}_0 \, .
\end{equation}
In Eq.~\eqref{eq:TVarOperator} the second term vanishes and the variance of
the time coordinate operator reads
\begin{equation}
\label{eq:2TVarOperator}
\Var{\hat{t};g(t,r)} = \sigma_t \,  r^2 \, .
\end{equation}
Similarly, for the radial coordinate operator $\hat{r}$ we get
\begin{equation}
\label{eq:RVarOperator}
\Aver{\hat{r}^2;g(t,r)}
= \frac{\Aver{\check{q}^2}_0}{\Aver{\check{q}}_0^2}\, r^2 \, ,
\end{equation}
so that the variance of $\hat{r}$ becomes
\begin{equation}
\label{eq:2RVarOperator}
\Var{\hat{r};g(t,r)} = \sigma_r \,  r^2 \, ,
\end{equation}
where
\begin{equation}
\label{eq:3RVarOperator}
\sigma_r :=
\frac{\Aver{\check{q}^2}_0- \Aver{\check{q}}_0^2}{
 \Aver{\check{q}}_0^2 } \, .
\end{equation}
In both cases the standard deviation from the expectation value (square root of
the variance) is proportional to the radius $r$. The coefficients in
\eqref{eq:3TVarOperator} and \eqref{eq:3RVarOperator} depend only on the
fiducial vector $\Phi_0(x)$.
One can see that while approaching to classical singularity $r \to 0$, the
  quantum radial observable behaves as the classical one because its expectation
  value goes to zero and its variance also goes to zero. However, the ratio of
  the standard deviation from the expectation value to the expextation value of
  $\hat{r}$ is constant. This suggests an existence of non-zero relative
  fluctuations of the radial coordinate even at singularity. Such fluctuations
  can be a germ which leads to larger fluctuations of other quantum observables,
  like spacetime invariants, and finally to avoiding the singularity in the
  Schwarzschild spacetime. 

To find the lowest bound of the product $\sigma_t \sigma_r$ one can use the
Heisenberg type uncertainty principle in the form proposed by Robertson
\cite{Robertson1929}. In this case we get
\begin{equation}
\label{eq:LBsigmatr}
\Var{\hat{t};g(t,r)}\,\Var{\hat{r};g(t,r)} \ge
\frac{\Aver{i[\check{p},\check{q}]}_0^2}{4 \Aver{\check{q}}_0^2}\,  r^4 \, ,
\end{equation}
which gives  the required lowest bound for product of both smearing coefficients
\begin{equation}
\label{eq:2LBsigmatr}
\sigma_t \,\sigma_r \ge
\frac{\Aver{i[\check{p},\check{q}]}_0^2}{4 \Aver{\check{q}}_0^2} \, .
\end{equation}
As an example we give values of the above constants for some particular fiducial
vectors. Let us take
\begin{equation}
\label{eq:fidvecEx}
\Phi_0(x)=\frac{1}{\sqrt{(2n-1)!}}x^ne^{-\frac{x}{2}} \, ,
\end{equation}
 where $n > 1$ is a natural number selected to ensure the convergence
 properties.

 One easily gets
\begin{eqnarray}
&&\sigma_t=\frac{2n-1}{2n-2} \label{eq:sigmatEx} \, ,\\
&&\sigma_r=\frac{1}{2n-2}  \label{eq:sigmarEx} \, .
\end{eqnarray}
Thus, the inequality (\ref{eq:2LBsigmatr}) reads
\begin{equation}
\sigma_t\,\sigma_r \ge
\frac{1}{4(2n-1)^2} \, .
\end{equation}


\section{Quantization of the Kretschmann scalar}

Observables which characterize the behaviour of the spacetime at a given
  spacetime point are the curvature invariants. In our case the most important is
  the Kretschmann scalar \eqref{Kret} .  The classical Kretschmann invariant
diverges as $r \rightarrow 0$.  Does this singularity survive quantization?  Is
the expectation value of the Kretschmann operator $\hat{\mathcal{K}}$ regular
across the configuration space $T$? What is the quantum smearing of
$\hat{\mathcal{K}}$?  These are the issues to be addressed in this section.

Using our quantization rules \eqref{eq:3AffQuantization}, the quantum
Kretschmann observable can be written as
\begin{equation}
\label{eq:QuantKret}
\hat{\mathcal{K}}=
48 M^2 \Aver{\check{q}}_0^6  \frac{1}{A_{\Phi_0}} \int_{\Aff} d\mu(p,q)
\Ket{g(p,q)} \frac{1}{q^6} \Bra{g(p,q)} \, .
\end{equation}
%
\subsection{Eigenproblem for $\hat{\mathcal{K}}$ operator}

As the first step, let us consider the eigenproblem of the operator
$\hat{\mathcal{K}}$ which allows to establish eigenfuctions (or rather
generalized eigenfuctions) and spectrum of the Kretschmann operator
\begin{equation}
\label{eq: KretschmannOp}
\int_{\RNumb_+}
d\nu(y)\;\mathbf{K}_\mathcal{K}(x,y)\;\psi^{(\mathcal{K})}_{k}(y)=
k\,\psi^{(\mathcal{K})}_{k}(x) \,,
\end{equation}
written in terms of the integral kernel
\begin{eqnarray}
\label{eq: KretschmannKern}
&&\mathbf{K}_\mathcal{K}(x,y)=\Bra{x}\hat{\mathcal{K}}\Ket{y}=
\frac{\Aver{\check{q}}_0^6}{A_{\Phi_0}} \int_{\Aff}d\mu(p,q)\;
\BraKet{x}{g(p,q)}\;\frac{48M^2}{q^6}\;\BraKet{g(p,q)}{y}= \nonumber \\
&&
=\frac{48M^2}{A_{\Phi_0}} \Aver{\check{q}}_0^6
\left[\int_{\RNumb_+}\frac{dq}{q^8}\left|\Phi_0(q)\right|^2\right]\;
\delta(x-y) x^7= \mathcal{A}\; \delta(x-y) x^7 \, ,
\label{eq:KretConst}
\end{eqnarray}
where the coefficient
$\mathcal{A}=\frac{48M^2}{A_{\Phi_0}} \Aver{\check{q}}_0^6
\left[\int_{\RNumb_+}\frac{dq}{q^8}\left|\Phi_0(q)\right|^2\right]$.
It must be noticed that the condition $\mathcal{A}<\infty$ requires an
appropriate behavior of the fiducial vector at $x$ equal to zero and infinity.

Direct calculations lead to the following generalized eigenfunctions
\begin{equation}
\psi^{(\mathcal{K})}_{k}(x)=
\delta\left(x^6-\frac{k}{\mathcal{A}}\right), \quad 0<k < \infty \,,
\end{equation}
and the positive spectrum $0<k < \infty$ of the  Kretschmann operator.

For further interpretation it is useful to calculate a form of these solutions
as functions of the affine group elements. In the standard parametrization
$(p,q)$ the above states can be written as
\begin{equation}
\label{eq: KretschmannPQ}
\psi^{(\mathcal{K})}_{k}(p,q)=\frac{1}{6}
\left(\frac{\mathcal{A}}{k}\right)^{\frac{5}{6}}
\exp\left[i\sqrt[6]{\frac{k}{\mathcal{A}}} p \right]\;\Phi_0^\star
\left(q \sqrt[6]{\frac{k}{\mathcal{A}}}\right) \, .
\end{equation}
It is obtained due to the useful transformation formula
\begin{equation}
\BraKet{g(p,q)}{f}
= \int_{\RNumb_+}d\nu(x)\;e^{-i px}\Phi_0^\star (qx) f(x) \, .
\end{equation}
According to general quantum rules one can expect that
$|\psi^{(\mathcal{K})}_{k}(t,r)|^2$ is related to density probability (in this
case it cannot be normalized) of finding the Schwarzschild spacetime in the
Kretschmann observable eigenstate if this physical system is in the coherent
state. As one can see this density probability is independent of $t$ and depends
only on the explicit form of the fiducial vector.

An important information implied by the eigenproblem solution of
$\hat{\mathcal{K}}$  is that the quantum Kretschmann scalar
can be potentially infinite because its spectrum is not bounded from above.

\subsection{Expectation value for the $\hat{\mathcal{K}}$ operator}
\label{subs:ExK}

Expectation values which give a link between quantum theory and observed values
of quantum observables are state dependent.
This feature is related to an important question about quantum states of our
  physical system. As we mentioned earlier, the fundamental observables
  $\hat{t}$ and $\hat{r}$ do not commute, but classically they are good
  observables of our quantum system so that the Schwarzschild spacetime cannot be
   in any common eigenstate of $\hat{t}$ and  $\hat{r}$. In
  fact, it is a consequence of the Heisenberg uncertainty principle. In this
  context we need to check if the expectation values of the operator
  $\hat{\mathcal{K}}$, determined  witin the coherent states representing elementary states
  of the spacetime, behave like the classical Kretschmann scalar.

Using the formula \eqref{eq:funq} from the appendix \ref{app:0}, one gets simple
general expression for the expectation value of the Kretschmann operator
\begin{equation}
\label{eq:AverK}
\Bra{\Psi}\hat{\mathcal{K}}\Ket{\Psi} \equiv \Aver{\hat{\mathcal{K}}; \Psi}
=\mathcal{A} \int_{\RNumb_+} dx\, x^6 |\Psi(x)|^2 \, .
\end{equation}
It turns out that the classical form of the Kretschmann scalar is
  proportional to the expectation value of the Kretschmann operator calculated
  within the coherent states $\Ket{g(t,r)}$ fulfilling the consistency
  conditions
\begin{equation}
\label{eq:ExpectValKret}
\Aver{\hat{\mathcal{K}}; g(t,r)}
=48M^2 \frac{\Aver{(q^{-6})\check{\phantom{w}}}_0}{\Aver{\check{q}}^{-6}_0}
\frac{1}{r^6}  \, .
\end{equation}
Therefore, the mean value $\Aver{\hat{\mathcal{K}}; g(t,r)}$ has formally the
singularity at $r=0$, as in the classical case.

 However, to determine its behaviour in quantum case fully we have to
  calculate its variance.  Applying \eqref{eq:ffAverGpar} to the operator
  \eqref{eq:QuantKret} gives
\begin{equation}
\label{eq:2ExpectValKret}
\Aver{\hat{\mathcal{K}}^2;g(t,r)}=
(48M^2)^2
\frac{\Aver{((q^{-6})\check{\phantom{w}})^2}_0}{\Aver{\check{q}}^{-12}_0}
\frac{1}{r^{12}} \, .
\end{equation}
Combining the expressions \eqref{eq:ExpectValKret} and \eqref{eq:2ExpectValKret}
we get the required variance of the Kretschmann operator within the coherent
states
\begin{equation}
\label{eq:VarianceKret}
\Var{\hat{\mathcal{K}};g(t,r)}=
(48M^2)^2
\left(\Aver{((q^{-6})\check{\phantom{w}})^2}_0
- \Aver{(q^{-6})\check{\phantom{w}}}_0^2 \right)
\Aver{\check{q}}^{12}_0 \; \frac{1}{r^{12}} \, .
\end{equation}
The variance \eqref{eq:VarianceKret} tends also to infinity as $r$ approaches
zero.  However, the ratio of the expectation value
$\Aver{\hat{\mathcal{K}}; g(t,r)}$ and the standard deviation
$\sqrt{\Var{\hat{\mathcal{K}};g(t,r)}}$ is independent on $r$ and $t$
\begin{equation}
\label{eq:AverToVarKret}
s= \frac{\Aver{(q^{-6})\check{\phantom{w}}}_0}{ %
  \sqrt{\Aver{((q^{-6})\check{\phantom{w}})^2}_0
-\Aver{(q^{-6})\check{\phantom{w}}}_0^2} } \, ,
\end{equation}
i.e., both, the expectation value of $\hat{\mathcal{K}}$  and its standard
deviation are proportional.
\begin{figure}
\begin{center}
\includegraphics[width=0.9\textwidth]{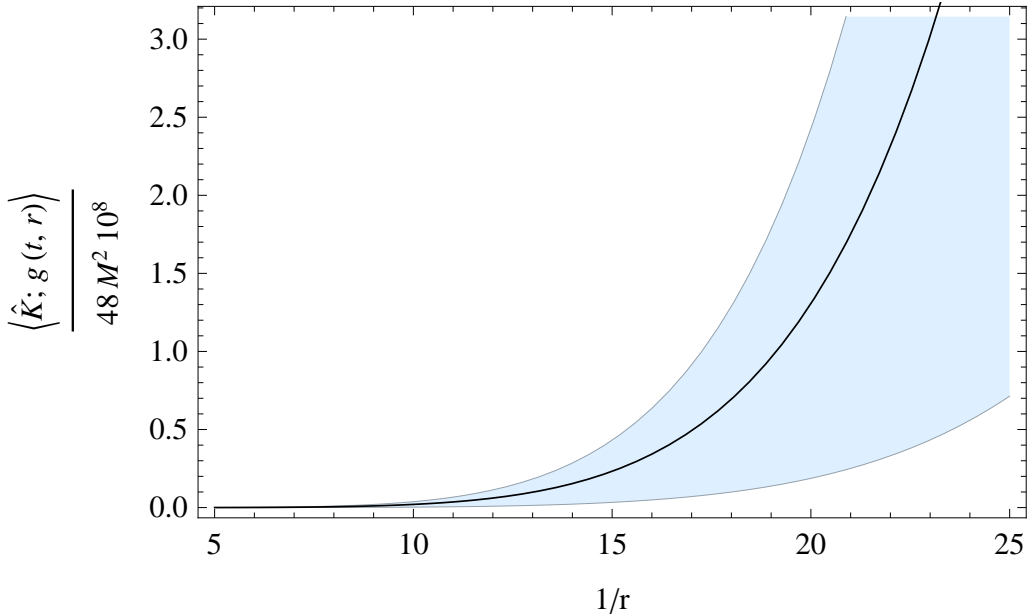}
\caption{\label{plot:KExpVar} The $1/r$ dependence of the expectation value of
 the Kretschmann operator $\Aver{\hat{\mathcal{K}}; g(t,r)}$ defined by
  (\ref{eq:ExpectValKret}).  The blue area defines the points for which
  distance from expected value is smaller than
  $\sqrt{\Var{\hat{\mathcal{K}};g(t,r)}}$ (the distance is counted along fixed
  $1/r$ line). The fiducial vector is taken as in (\ref{eq:fidvecEx}) with
  $n=25$. }
\end{center}
\end{figure}
This behavior of the variance protects the mean value of the quantum Kretschmann
observable within the coherent states to be singular. The operator
$\hat{\mathcal{K}}$ represents a well behaving smeared observable which is
completely undetermined at the classical singularity $r=0$, see
Fig.\ref{plot:KExpVar}.   Fluctuations of the Kretschmann quantum observable
  grow to infinity.

This is a novel mechanism which allows to omit singularity
after quantization of classical variables.

Above, our new mechanism was checked only on the fundamental set of states,
  i.e., for the affine coherent states. In the appendix \ref{app:2} we show
that the expectation values and variances of the Kretschmann operator within the
dense set of states
\begin{equation}\label{dense}
\Psi_n(x)= N x^n\exp\left[i\tau_0 x-\frac{\gamma^2 x^2}{2}\right],
\end{equation}
where $N^2=2\gamma^n/(n-1)!$ and $n=1,2,\dots$, behaves exactly in the same
way as it was obtained for the affine coherent states.

\section{Conclusions}

The extension of the configuration space to include temporal variable at the
same footing as spatial variables is the novelty in the programme of
quantization of gravity.  In this paper we have used this idea to address the
issue of the fate of the naked gravitational singularity of the Schwarzschild
spacetime at the quantum level.  Quantization of the time variable has enabled
resolving the singularity problem. The above idea seems to be fruitful and worth
of being applied to more realistic models of spacetime with naked singularities
like the ones considered in a series of papers by Pankaj Joshi and his
collaborators (see, e.g., \cite{PJ1,PJ2,PJ3,PJ4,PJ5,PJ6,PJ7,PJ8}
and references therein).

If isolated objects with naked singularities do occur in the real world, their
examinations may bring highly valuable data to be used in the construction of
quantum gravity.  It is so because the isolated objects with covered
singularities, i.e. black holes, may have screened some essential quantum
gravity data due to the presence of horizons.

The solution of the eigenproblem for the Kretschmann operator shows that the
spectrum is bounded from below and unbounded from above. The latter seems to
lead to an embarrassment, but further examinations in the context of expectation
value and the variance of the Kretschmann operator indicate the resolution of
this difficulty.

Making use of the affine coherent states quantization, we have found that the
expectation value of the Kretschmann operator $\hat{\mathcal{K}}$ is singular
and behaves like $1/r^6$ as in the classical case. However, its variance behaves
like $1/r^{12}$. One can say that quantization smears the singularity, avoiding
its localization in the region of the configuration space including the
singularity.  In addition, since the variance not only does not vanish but
diverges as $r \rightarrow 0$, the state corresponding to $r=0$ cannot be any
eigenstate of the operator $\hat{\mathcal{K}}$, which is suggested by the
property \eqref{lem}. Thus, the system cannot occupy the state corresponding to
the gravitational singularity. One can say that probability of finding our
system in the singular state is equal to zero.

The above result, carried out for the affine coherent states, has been confirmed
in App. \ref{app:2} for any vector of the carrier space $L^2(\RNumb_+,d\nu(x))$.
This proves the generality of our singularity avoiding mechanism.  Our conclusion
seems to be true for any quantum state of the system under consideration.

The issue of possible resolution of the singularity problem of the Schwarzschild
black hole ($M > 0$) at quantum level, has been addressed in several papers (see, e.g.,
\cite{Blan1,Abhay,Lisa} and references therein).  It is based on the isometry of
the interior of the black hole with the vacuum Kantowski-Sachs spacetime.  An
interesting approach is presented in \cite{Blan1}.  The corrections to the
Raychaudhuri equation in the interior of the Schwarzschild black hole derived
from loop quantum gravity (LQG) has been examined.  The resulting effective
equation implies the defocusing of geodesics which prevents the formation of
conjugate points so that leads to the resolution of the singularity problem. In
\cite{Abhay} the Kruskal-Szekeres coordinates \cite{Piotr} are applied. Quantum
corrections of LQG are used to resolve the singularity problem, and the
resulting quantum extension of spacetime has interesting features. An effective
LQG model of the Schwarzschild black hole interior based on Thiemann's
identities is proposed in \cite{Lisa}. The effective dynamics leads to the
resolution of the classical singularity. A spherically symmetric vacuum gravity
is quantized using LQG techniques in \cite{Jorge}. Dirac's quantization
procedure leads to the resolution of the singularity of the classical theory
inside black holes. The loop quantization of the model of Schwarzschild interior
coupled to a massless scalar field has been studied \cite{Ma}. Obtained results
indicates the existence of a non-vanishing minimal mass of that black hole,
which implies the existence of some black hole remnants after the Hawking
evaporation.

An extension of the present paper to the case of the Schwarzschild black hole
is straightforward. It will be considered in the context of quantization of the
Lemaître-Tolman-Bondi model of isolated object, with naked or covered singularity,
in the near future \cite{Janek}.

\acknowledgments We would like to thank  Jan Ostrowski for helpful  discussions.

\appendix

\section{Some remarks about  calculations}
\label{app:0}

According to methodology of integral quantization,one can see that for every
classical observable $f(t,r)$ the corresponding quantized operator $\hat{f}$
(see \eqref{eq:3AffQuantization}) is symmetric because its quadratic form
\begin{equation}
\label{eq:4AffQuantization}
\Bra{\Psi}\hat{f}\Ket{\Psi}= \frac{1}{A_{\Phi_0}} \int_{\Aff} d\mu(p,q)
|\BraKet{g(p,q)}{\Psi}|^2
f\left(\chi^{-1}\left(p-\frac{a}{b}q,\frac{q}{b} \right)\right)
\end{equation}
is real for $\Psi$ belonging to the domain of the operator $\hat{f}$.
The operator $\hat{f}$ can be bounded by the following expression
\begin{eqnarray}
\label{app0:BoundA}
&& \Vert \hat{f} \Ket{h} \Vert^2
= \Big\vert
\frac{1}{A_{\Phi_0}^2}
\int_{\Aff} d\mu(p_1,q_1) \int_{\Aff} d\mu(p_2,q_2)  \nonumber \\
&& f\left(\chi^{-1}\left(p_1-\frac{a}{b}q_1,\frac{q_1}{b} \right)\right)
f\left(\chi^{-1}\left(p_2-\frac{a}{b}q_2,\frac{q_2}{b} \right)\right)
\BraKet{h}{g(p_1,q_1)} \BraKet{g(p_1,q_1)}{g(p_2,q_2)}\BraKet{g(p_2,q_2)}{h}
\Big\vert \nonumber \\
&& \leq \frac{1}{A_{\Phi_0}^2}
\int_{\Aff} d\mu(p_1,q_1) \int_{\Aff} d\mu(p_2,q_2)
\Big\vert f\left(\chi^{-1}\left(p_1-\frac{a}{b}q_1,\frac{q_1}{b} \right)\right)
\nonumber \\
&& f\left(\chi^{-1}\left(p_2-\frac{a}{b}q_2,\frac{q_2}{b} \right)\right)
\BraKet{h}{g(p_1,q_1)} \BraKet{g(p_1,q_1)}{g(p_2,q_2)}\BraKet{g(p_2,q_2)}{h}
\Big\vert \nonumber \\
&& \leq
\Big[\int_{\Aff} d\mu (p_1,q_1) \int_{\Aff} d\mu (p_2,q_2)
\Big\vert
f\left(\chi^{-1}\left(p_1-\frac{a}{b}q_1,\frac{q_1}{b} \right)\right)
\BraKet{g(p_1,q_1)}{g(p_2,q_2)} \nonumber \\
&& f\left(\chi^{-1}\left(p_2-\frac{a}{b}q_2,\frac{q_2}{b} \right)\right)
 \Big\vert \, \Big]
{\Vert \, \Ket{h} \,\Vert}^2 \, ,
\end{eqnarray}
for all $\Ket{h}$ in the domain of the operator $\hat{f}$. The last step is
obtained by making use of the Schwartz inequality
$|\BraKet{g(p,q)}{h}| \leq \Vert \Ket{h} \Vert$. If the above integral contained
in square bracket is finite the operator $\hat{f}$ is continuous in
$L^2(\RNumb_+,d\nu(x))$, i.e., $\hat{f}$ is a self-adjoint operator.

However, in practice, even the elementary observables $\hat{t}$ and $\hat{r}$
are unbounded operators and require more careful procedures of extension their
domains.  Matrix elements of operators are crucial expressions required in
quantum calculations. They can be used to extend such operators by
symmetrization of their matrix elements
\begin{equation}
\label{app0:1OpSymmetrization}
\Bra{\psi_2}\hat{A}\Ket{\psi_1}_{sym} := \frac{1}{2} \left(
\BraKet{\psi_2}{\hat{A}\psi_1}+ \BraKet{\psi_1}{\hat{A}\psi_2}^\star \right) \, .
\end{equation}
For any symmetric operator $\hat{A}$  the following identity hold
\begin{equation}
\label{app0:2OpSymmetrization}
\Bra{\psi_2}\hat{A}\Ket{\psi_1}_{sym} = \Bra{\psi_2}\hat{A}\Ket{\psi_1} \ ,
\end{equation}
for $\psi_1$ and $\psi_2$ in the domain of $\hat{A}$, however, in other cases
this equality can be broken and then the symmetrization
\eqref{app0:1OpSymmetrization} is useful.

Let us assume that $A(p',q')$, where $p'=p'(p,q)$ and
$q'=q'(p,q)$ are real functions. A typical matrix elements are of the following
form
\begin{equation}
\label{app0:1MatElem}
\Bra{\psi_2}\hat{A}\Ket{\psi_1} =
\frac{1}{A_{\Phi_0}}
\int_{\Aff} d\mu(p,q) \BraKet{\psi_2}{g(p,q)} A(p',q') \BraKet{g(p,q)}{\psi_1}\,.
\end{equation}
Calculating in the space $L^2(\RNumb_+,d\nu(x))$ we allow for changing of
integration order
\begin{eqnarray}
\label{app0:2MatElem}
&&\Bra{\psi_2}\hat{A}\Ket{\psi_1} =
\int_{\RNumb_+} d\nu(x_2) \int_{\RNumb_+} d\nu(x_1)  \nonumber \\
&& \psi_2(x_2)^\star \left\{
\frac{1}{A_{\Phi_0}} \int_{\Aff} d\mu(p,q)
e^{ip(x_2-x_1)} A(p',q') \Phi_0(qx_2) \Phi_0(qx_1)^\star \right\} \psi_1(x_1)
\nonumber \\
&& = \int_{\RNumb} dx_2 \int_{\RNumb} dx_1
\theta(x_2)\frac{1}{x_2}\psi_2(x_2)^\star \nonumber \\
&& \left\{
\frac{1}{A_{\Phi_0}} \int_{\Aff} d\mu(p,q)
e^{ip(x_2-x_1)} A(p',q') \Phi_0(qx_2) \Phi_0(qx_1)^\star
\right\}
\theta(x_1)\frac{1}{x_1}\psi_1(x_1)
\end{eqnarray}
and we extend integration over $x$ variables on the whole real axis adding the
Heaviside function $\theta(x)$. This is useful for regularization of integrals,
if needed, in the spirit of distribution theory.

As an example, let us consider the operator and its matrix element between the
position eigenstate $\Ket{x}$ and any arbitrary vector in the
$L^2(\RNumb_+,d\nu(x))$ space
\begin{equation}
\label{app1:OperatorP}
\Bra{x}\hat{p}\Ket{\psi} = \theta(x) \frac{1}{2\pi A_\phi}
\int_{\RNumb} dp \int_{\RNumb_+} \frac{dq}{q^2}
\BraKet{x}{g(p,q)} p \BraKet{g(p,q)}{\psi} \ .
\end{equation}
Using the explicit form of the scalar products we get
\begin{equation}
\label{app1:2OperatorP}
\Bra{x}\hat{p}\Ket{\psi} = \frac{1}{A_\phi}
\theta(x) \int_{\RNumb_+} \frac{dq}{q^2}
\int_{\RNumb} dy
\left(\int_{\RNumb} dp\, p e^{ip(x-y)} \right)
\Phi_0(qx) \Phi_0(qy)^\star  \theta(y) \frac{1}{y}\psi(y) \ .
\end{equation}
To regularize the integral over $p$, we use the known formula
\begin{equation}
\label{app1:RegIntegralP}
\int_{\RNumb} dp\, p\, e^{ip(x-y)}=-i 2\pi \delta'(x-y) \ ,
\end{equation}
where prime denotes distributional derivative of the Dirac delta.
After using this expression and definition of $\delta'$ one gets
\begin{eqnarray}
\label{app1:3OperatorP}
&& \Bra{x}\hat{p}\Ket{\psi} = \frac{-i}{A_\phi}
\theta(x) \int_{\RNumb_+} \frac{dq}{q^2} \Phi_0(qx)
\frac{\partial}{\partial x}
\left[\theta(x) \Phi_0(qx)^\star \frac{1}{x}\psi(x)\right] \nonumber \\
&& = -i \theta(x) \delta(x) \psi(x)
+ \theta(x) \left(-i\frac{\partial}{\partial x}+\frac{i}{2x} \right) \psi(x)\,.
\end{eqnarray}
Note, that in this case the position state $\Ket{x}$ does not belong to the
domain of the operator $\hat{p}$ and using it can require symmetrization.

This formula allows to write more general matrix element
\begin{eqnarray}
\label{app1:4OperatorP}
&& \Bra{\psi_2}\hat{p}\Ket{\psi_1} =
-i \theta(0) \lim_{x \to 0^+} \frac{\psi_1(x)^\star\psi_2(x)}{x} \nonumber \\
&& + \int_{\RNumb_+} d\nu(x) \psi_2(x)^\star
\left(-i\frac{\partial}{\partial x}+\frac{i}{2x} \right) \psi_1(x) \, ,
\end{eqnarray}
which after symmetrization can be rewritten as
\begin{eqnarray}
\label{app1:5OperatorP}
&& \Bra{\psi_2}\hat{p}\Ket{\psi_1}_{sym} = \theta(0)
\lim_{x \to 0^+} \frac{\mathrm{Im}( \psi_2(x)^\star \psi_1(x))}{x}
\nonumber \\
&& + \frac{(-i)}{2}\int_{\RNumb_+} d\nu(x)
\left(
\psi_2(x)^\star \frac{\partial \psi_1}{\partial x}
- \frac{\partial \psi_2}{\partial x}^\star \psi_1(x)
\right) \ .
\end{eqnarray}
Note that for the real functions $\psi$ the expectation value
$\Bra{\psi}\hat{p}\Ket{\psi}_{sym}=0$. Even more general, the expectation value
$\Bra{\psi}\hat{p}\Ket{\psi}_{sym}=0$ for
$\mathrm{Im}\left(\psi(x)^\star \frac{\partial \psi}{x} \right)=0$.

An interesting quantity is the expectation value of the $\hat{p}$ operator in
the gaussian wave packet $\Psi^{(t)}(x)$ constructed from generalized eigenstates
$\psi^{(t)}_{\tau}(x)=\sqrt{x} e^{i\tau x}$ of the $\hat{p}$ operator
\begin{equation}
\label{app1:GaussianPackettime}
\Psi^{(t)}(x)=N_t \int_{\RNumb} d\tau f^{(t)}(\tau) \sqrt{x} e^{i\tau x}
= N_t \sqrt{x} \exp{(i\tau_0 x)} \exp{\left[-\frac{1}{2} \gamma_t^2 x^2 \right]}
\ ,
\end{equation}
where
\begin{equation}
\label{app1:2GaussianPackettime}
f^{(t)}(\tau)= \frac{1}{\sqrt{2\pi}\gamma_t}
\exp{\left[-\frac{(\tau-\tau_0)^2}{2\gamma_t^2}\right]} \ .
\end{equation}
The normalization coefficient is equal to $N_t^2=2\gamma_t/\sqrt{\pi}$.  Making
use of the formula \eqref{app1:4OperatorP} and $\theta(0)=1/2$ the required
average value is
\begin{eqnarray}
\label{app1:3GaussianPackettime}
\Bra{\Psi^{(t)}(x)}\hat{p}\Ket{\Psi^{(t)}(x)} = \tau_0 \ ,
\end{eqnarray}
as it is expected. The same result one obtains from \eqref{app1:5OperatorP}.

Using the methods of this section, a rather general form of matrix elements can
be obtained if the classical observable is dependent only on $q=r$ variable. In
this case we need to quantize the function $f(r)$
\begin{equation}
\label{eq:funq}
\Bra{\Psi} \hat{f} \Ket{\Psi}
=\frac{1}{A_{\Phi_0}} \int_{\Aff} d\mu(p,g) f(q) |\BraKet{g(p,q)}{\Psi}|^2
= \frac{1}{A_{\Phi_0}} \int_{\RNumb_+} dx\, \frac{|\Psi(x)|^2}{x^2}
\int_{\RNumb_+} \frac{dq}{q^2} f(q) |\Phi_0(qx)|^2
\end{equation}
%

\section{Eigensolutions for elementary  operators}
\label{app:1}

Because of the importance of the operators $\hat{t}$ and $\hat{r}$,
it is interesting to find the eigensolutions of these operators.
We will compute them without assuming any special form of fiducial vectors,
the only assumption  is that a fiducial vector is a real function.

In this case, the constants  $a$ and $b$ are as follows
\begin{eqnarray}
&&a=\Aver{\check{p}}_0=0  \label{eq:aConsExample} \,, \\
&&b=\Aver{\check{q}}_0=1  \label{eq:bConsExample} \  .
\end{eqnarray}
They are calculated in App. \ref{app:4}.

\subsection{Eigenproblem for $\hat{t}$ operator}

It is easy to show that eigenfunctions of the differential part of the operator
(\ref{app1:3OperatorP})
\begin{equation}
\label{app1:RedTOperator}
 \left(-i\frac{\partial}{\partial x}+\frac{i}{2x} \right)\psi^{(t)}_\tau(x)=\tau \psi^{(t)}_\tau(x)
\end{equation}
are equal to
\begin{equation}
\label{app1:RedTSolution}
\psi^{(t)}_\tau(x)=\sqrt{x}e^{i\tau x} \ .
\end{equation}
This implies the following matrix elements of the operator $\hat{t}$
\begin{eqnarray}
&&\Bra{\psi}\hat{t}\Ket{\psi^{(t)}_\tau}=
-i\theta(0)\lim_{x\rightarrow 0^+}\frac{\psi^\star(x)}{\sqrt{x}}+
\tau\BraKet{\psi}{\psi^{(t)}_\tau} \ .
\end{eqnarray}
Because every function $\psi \in L^2(\RNumb_+,d\nu(x))$ has to converge to 0 as
fast as $x$, or faster, when $x$ is going to $0^+$, the condition
$\lim_{x\rightarrow 0^+}\psi^\star(x)/\sqrt{x}=0$ is fulfilled for such functions
and the solutions (\ref{app1:RedTSolution}) are generalized eigensolutions (weak
solutions) of the operator $\hat{t}$.

It is also possible to solve the eigenequation for $\hat{t}$ by using
integral kernel as it was done for the operator $\hat{\mathcal{K}}$.
In this case the integral kernel is equal to
\begin{eqnarray}
&&\mathbf{K}_t(x,y)=\Bra{x}\hat{t}\Ket{y}
=\frac{1}{A_{\Phi_0}}\int_{\Aff} d\mu(pq)
\BraKet{x}{g(p,q)}\left(p-\frac{a}{b}q\right)\BraKet{g(p,q)}{y}=
\nonumber \\
&& =\frac{(-i)}{A_{\Phi_0}}\delta'(x-y)
\int_{\RNumb_+}\frac{dq}{q^2}\Phi_0(xq)\Phi^\star_0(qy)  . \label{eq:KertExample}
\end{eqnarray}
Using this method one has to extend the integration over $y$ to the whole real
axis by introducing under integral the Heaviside function, as it is shown in the
appendix \ref{app:0}.

It is interesting to show a form of $\psi^{(t)}_{\tau}(x)$ as a function of
$(p,q)$ variables. For this purpose one needs to use an explicit form of a
fiducial vector. For example, let us assume
$\Phi_0(x)=\frac{1}{\sqrt{(2n-1)!}}x^ne^{-\frac{x}{2}}$ and then
\begin{equation}
\psi^{(t)}_{\tau}(p,q)=\int_{\RNumb_+}
d\nu(x)\;\BraKet{g(p,q)}{x}\psi^{(t)}_{\tau}(x)
=\frac{\Gamma\left(n+\frac{1}{2}\right)}{\sqrt{(2n-1)!}}
\frac{q^n}{\left(\frac{q}{2}+i(p+\tau)\right)^{n+\frac{1}{2}}} \, .
\end{equation}
Obviously, $\psi^{(t)}_{\tau}$ does not belong to $\mathcal{H}_x$ because it is
not a square integrable function with the measure $d\nu(x)$ on $\RNumb_+$.
%

\subsection{Eigenproblem for $\hat{r}$ operator}

Now, let us examine the eigensolutions of the operator $\hat{r}$.
In this case the assumption about reality of the fiducial vector is not needed.
\begin{equation}
\int_{\RNumb_+}
d\nu(y)\;\mathbf{K}_r(x,y)\;\psi^{(r)}_{s}(y)=s\,\psi^{(r)}_{s}(x)\,,
\end{equation}
where
\begin{eqnarray}
&& \mathbf{K}_r(x,y)=\Bra{x}\hat{r}\Ket{y}=
\frac{1}{A_{\Phi_0}}\int_{\Aff}d\mu(p,q)\;
\BraKet{x}{g(p,q)}\;\frac{q}{b}\;\BraKet{g(p,q)}{y}= \nonumber \\
&&=\frac{1}{A_{\Phi_0}}\delta(x-y)
\, .
\end{eqnarray}
Eigenfunctions of $\hat{r}$ are Dirac delta type functions
\begin{eqnarray}\label{eigenr}
&&\psi^{(r)}_{s}(x)
=\delta\left(x-\frac{1}{A_{\Phi_0} s}\right)\, .
\end{eqnarray}
The form of these solutions as a functions of $(p,q)$ reads
\begin{equation}
\psi_{s}(p,q)=\int_0^\infty d\nu(x)
\BraKet{g(p,q)}{x}\, \psi^{(r)}_{s}(x)
= A_{\Phi_0}\;s\,
\exp\left[-\frac{ip}{A_{\Phi_0} s}\right]\;
\Phi^\star_0\left(\frac{q}{A_{\Phi_0} s}\right)\, .
\end{equation}
Obviously, $\psi^{(r)}_{s}$ does not belong to $\mathcal{H}_x$ because it is not
a square integrable function with the measure $d\nu(x)$ on $\RNumb_+$.

\section{Expectation values of the Kretschmann operator \\within
  a basis in the Hilbert space $L^2(\RNumb_+,d\nu(x))$}
\label{app:2}

In this appendix  we present a derivation of expectation values and
variances of the operators $\hat{t}$, $\hat{r}$ and $\hat{\mathcal{K}}$  within a class of
quantum states furnishing a basis in the Hilbert space $L^2(\RNumb_+,d\nu(x))$.

Let us consider  quantum states similar to the wave packets
$\Psi^{(t)}(x)$ defined by \eqref{app1:GaussianPackettime}, where the only
modification is in $x$ dependence of the eigenfunctions $\psi^{(t)}_\tau$
\begin{equation}
\label{eq:DensSetPsin}
\Psi_n(x)=N\int_\RNumb d\tau f^{(t)}(\tau)x^n e^{i\tau x}=
N x^n\exp\left[i\tau_0 x-\frac{\gamma^2 x^2}{2}\right]\, ,
\end{equation}
where $n = 1,2, \ldots$, the $f^{(t)}(\tau)$ is a Gaussian distribution, and where $N^2=2\gamma^n/(n-1)!$.

The expectation values of the operators $\hat{t}$, $\hat{r}$,
$\hat{\mathcal{K}}$ in the states $\Psi_n$ are as follows
\begin{eqnarray}
\label{eq:PsiNBeg}
&&\Aver{\hat{t}; \Psi_n}=\tau_0 \,, \\
&&\Aver{\hat{r}; \Psi_n}
=\frac{1}{A_\Phi}\frac{\Gamma\left(n-\frac{1}{2}\right)}{(n-1)!}\;\gamma \, ,\\
&&\Aver{\hat{\mathcal{K}}; \Psi_n}
=\mathcal{A}\frac{(n+2)!}{(n-1)!}\;\frac{1}{\gamma^6}\, ,
\end{eqnarray}
and the corresponding variances are
\begin{eqnarray}
\label{eq:PsiNEnd}
&&\Var{\hat{t};\Psi_n}=\frac{4n-3}{4(n-1)}\gamma^2 \, ,\\
&&\Var{\hat{r};\Psi_n}=\frac{1}{A^2_\Phi} \left(\frac{1}{n-1}
- \frac{\Gamma\left(n-\frac{1}{2}\right)^2}{(n-1)!^2}\right)\;\gamma^2 \, ,\quad  n\geq 2 \\
&&\Var{\hat{\mathcal{K}};\Psi_n}= \mathcal{A}^2\left(\frac{(n+5)!}{(n-1)!}
-\frac{(n+2)!^2}{(n-1)!^2}\right)\;\frac{1}{\gamma^{12}} \, .
\end{eqnarray}
Average value and variance of $\hat{\mathcal{K}}$ and $\hat{r}$
are the same as in subsection \ref{subs:ExK}. The last statement is important
because we prove below that  the set of the functions $\Psi_n$ is dense in the
Hilbert space  $L^2(\RNumb_+,d\nu(x))$.

Namely, let us
consider the subset of functions $\Psi_n$ where index $n$ is odd.
By using linear combination of these functions one can build the set
of functions in the following form
\begin{equation}
\label{eq:1GaussBasisRplus}
l_k (x) :=\sqrt{2}\gamma \; x\;  L_k(\gamma^2 x^2)\;
\exp\left(i\tau_0 x -\frac{\gamma^2x^2}{2}\right)\, ,
\end{equation}
where $L_k(x)$ are the Laguerre polynomials. Calculating the scalar product,
after changing the variables $y=\gamma^2 x^2$, one gets
\begin{equation}
\label{eq:2GaussBasisRplus}
\BraKet{l_k}{l_m}
= \int_0^\infty d\nu(x) l_k(x)^\star l_m(x)
= \int_0^\infty dy L_k(y) L_m(y) e^{-y} =\delta_{km}.
\end{equation}
It means, the set of functions \eqref{eq:1GaussBasisRplus} form an orthonormal
basis in the Hilbert space $L^2(\RNumb_+,d\nu)$.  Therefore, the set of
functions $\Psi_n(x)$ must be dense in $L^2(\RNumb_+,d\nu(x))$ and every
function belonging to our Hilbert space can be expressed as a combination of the
functions $\Psi_n(x)$.

\section{Calculations of
  $\Aver{\check{p}}_0,\Aver{\check{q}}_0,\Aver{\check{p}^2}_0$, and
  $\Aver{\check{q}^2}_0$}
  \label{app:4}

In the following we present the calculation of component
parts which are needed for $\sigma_t$ and $\sigma_r$.
In the computation we use method described in the Appendix \ref{app:0}.
For shortness we use the following notation
$\BraKet{g(p_1,q_1)}{g(p_2,q_2)}=\BraKet{p_1,q_1}{p_2q_2}$ . 
%
Now, we calculate
\begin{eqnarray}
&&\Aver{\check{p}}_0=
\frac{1}{A_{\Phi_0}}\int_{\Aff}d\mu(p,q)\;\BraKet{0,1}{p,q}\;p\;\BraKet{p,q}{0,1}=
\\
&&
=\frac{i}{A_{\Phi_0}}
\left[
\int_{\RNumb_+}\frac{dq}{q^2}  \int_{\RNumb_+}\frac{dy}{y}
\left(\frac{\Phi^\star_0(y)}{y}\right)'\Phi_0(qy)
\Phi^\star_0(qy)\Phi_0(y)+\right.
\\
&&
+\int_{\RNumb_+}\frac{dq}{q}  \int_{\RNumb_+}\frac{dy}{y}
\frac{\Phi^\star_0(y)}{y}\Phi'_0(qy)
\Phi^\star_0(qy)\Phi_0(y)+
\\
&&
\left.
\int_{\RNumb_+}\frac{dq}{q}  \int_{\RNumb_+}\frac{dy}{y^2}\delta(y)\left|\Phi_0(y)\right|^2\left|\Phi_0(qy)\right|^2
\right] \,.
\end{eqnarray}
The second of these integrals can be turned to the form
$(-A_{\Phi_0})\int_{\RNumb_+} dy\;\Phi_0(y)^\star\Phi'_0(y)/y$.
The third one gives the same formula with the opposite sign.
The last one is equal to
$\lim_{x\rightarrow 0^+}\left|\Phi_0(y)\right|^2/y$.
Choosing a fiducial vector for which this limit is equal to zero, one gets
\begin{equation}
\Aver{\check{p}}_0=0 \, .
\label{appD:Averp}
\end{equation}

Now, we calculate further required coefficients of type $\Aver{\check{A}}_0$ :
\begin{eqnarray}
&&\Aver{\check{q}}_0=
\frac{1}{A_{\Phi_0}}\int_{\Aff}d\mu(p,q)
\BraKet{0,1}{p,q}\;q\;\BraKet{p,q}{0,1}  \\
&& =\frac{1}{A_{\Phi_0}}\int_{\RNumb_+}\frac{dq}{q}
\int_{\RNumb_+}\frac{dy}{y}\Phi^\star_0(y)\Phi_0(qy)\Phi^\star_0(qy)\Phi_0(y)=1 \, .
\end{eqnarray}
For variances we need  average values of squares of the operators
$\check{p}$ and $\check{q}$:
\begin{eqnarray}
&& \Aver{\check{p}^2}_0=\frac{1}{A_{\Phi_0}^2}
\int_{\Aff} d\mu(p_1,q_1)\int_{\Aff} d\mu(p_2,q_2)
\BraKet{0,1}{p_1,q_1}\;p_1\;
\BraKet{p_1,q_1}{p_2,q_2}\; \\
&& p_2\;\BraKet{p_2,q_2}{0,1} \\
&& =
\frac{(-1)}{A^2_{\Phi_0}}
\int_{\RNumb_+}\frac{dq_1}{q_1^2}
\int_{\RNumb_+} dx\;\frac{\Phi^\star_0(x)\Phi_0(q_1x)}{x}
\int_{\RNumb_+}
dy\frac{\Phi^\star_0(q_1y)}{y}\delta'(x-y)
\\
&& \left[
\int_{\RNumb_+}\frac{dq_2}{q_2^2}
\left(\int_{\RNumb_+} dz\;\delta'(z-y)\frac{\Phi^\star_0(q_2z)\Phi_0(z)}{z}\right)
\Phi^\star_0(yq_2)
\right] \, .
\end{eqnarray}
The integral in the square bracket is equal to:
\begin{eqnarray}
&&
A_{\Phi_0} y\left(\frac{\Phi_0(y)}{y}\right)'+\frac{\Phi_0(y)}{y}B
+A_{\Phi_0} \delta(y)\Phi_0(y) \, ,
\label{appD:P2Part1}
\end{eqnarray}
where $B=\int_{\RNumb_+} \frac{dq}{q}\Phi_0(q)\Phi'^\star_0(q)$. If $\Phi_0$ is a
real function, $B= A_{\Phi_0}/2$.

If one choose the fiducial vector in such a way that the limit
$\lim_{y\rightarrow 0^+}\frac{\Phi_0^\star(q_1y)\Phi_0(y)}{q}$
is equal to zero
for every $q_1$, then the integration of the part which includes
$\delta(y)$ is equal to zero. The similar situation one gets after
integration over $dx$ in the original integral.
If one take the fiducial vector fulfilling the conditions
$\lim_{y\rightarrow 0^+}\Phi_0(y)/y=0$ and $\lim_{y\rightarrow 0^+}(\Phi_0(y)/y)'<\infty$
one gets:
\begin{eqnarray}
&&\Aver{\check{p}^2}_0=\int_{\RNumb_+}\frac{dy}{y}
\left|
y\left(\frac{\Phi_0(y)}{y}\right)'+\frac{B}{A_\Phi}\frac{\Phi_0(y)}{y}
\right|^2 \, .
\end{eqnarray}

The last component reads
\begin{eqnarray}
&&
\Aver{\check{q}^2}_0
=\frac{1}{A^2_{\Phi_0}}\int_{\Aff} d\mu(p_1,q_1)\int_{\Aff} d\mu(p_2,q_2)
\BraKet{0,1}{p_1,q_1}\;q_1\;
\BraKet{p_1,q_1}{p_2,q_2}\; \\
&& q_2\;\BraKet{p_2,q_2}{0,1}
=\frac{1}{A_{\Phi_0}^2}\int_{\RNumb_+}\frac{dz}{z^3}\left|\Phi_0(z)\right|^2 \, .
\end{eqnarray}


\end{document}